\documentclass[jkps,reprint,fleqn,showpacs,showkeys]{revtex4}
\usepackage{graphicx}
\usepackage{amssymb}
\usepackage{amsmath}
\usepackage{bm}
\begin{document}
\setcounter{page}{0}
\title[]{Synchrotron X-ray Spectroscopy Study on the Valence State in $\alpha$- and $\beta$-YbAlB$_4$ at Low Temperatures 
and High Magnetic Fields}

\author{Y. H. \surname{Matsuda}}\thanks{Fax: +81-4-7136-3220}
\email{ymatsuda@issp.u-tokyo.ac.jp}
\author{T.  \surname{Nakamura}}
\author{K.  \surname{Kuga}}
\author{S.  \surname{Nakatsuji}}
\affiliation{Institute for Solid State Physics, University of Tokyo, Chiba 277-8581, Japan}

\author{S.  \surname{Michimura}}
\author{T.  \surname{Inami}}
\affiliation{Japan Atomic Energy Agency, Sayo, Hyogo 679-5148, Japan}

\author{N.  \surname{Kawamura}}
\author{M.  \surname{Mizumaki}}
\affiliation{SPring-8/Japan Synchrotron Radiation Research Institute,1-1-1 Kouto, Sayo, Hyogo 679-5198, Japan}

\date{\today}

\begin{abstract}
The valence state of Yb ions in $\beta$-YbAlB$_4$ and its polymorph $\alpha$-YbAlB$_4$ has been investigated 
by using X-ray absorption and emission spectroscopy in SPring-8 at temperatures from 2 to 280~K. 
The observed Yb valence is 2.78 $\pm$ 0.01 in $\beta$-YbAlB$_4$ at 2 K by using the X-ray emission spectroscopy.
The valence is found to gradually increase with increasing temperature toward the trivalent state, and the characteristic temperature 
of the valence fluctuation is expected to be about 290~K.
We also found a small increase in the Yb valence  ($\sim$ 0.002) by applying a magnetic field of 32~T at 40~K to $\beta$-YbAlB$_4$. 
\end{abstract}



\maketitle

\section{INTRODUCTION}

Valence fluctuation phenomena in heavy-fermion (HF) compounds have attracted much attention because the quantum critical point (QCP) 
of the valence transition can play an important role the novel phenomena such as the superconductivity and the non-Fermi liquid behaviors 
\cite{miyake, wata10}.
$\beta$-YbAlB$_4$ is theoretically proposed to be located very close to the QCP of the valence transition \cite{wata10}.
Actually, considerable valence fluctuation was observed experimentally \cite{ohkawa, matsumoto}.

Moreover, since the valence state directly reflects the degree of localization of the $f$-electrons in HF systems, the temperature and 
the magnetic field dependences of the valence are intriguing research subjects.  
There are several techniques to measure the valence state, e.g., photoemission spectroscopy, X-ray absorption spectroscopy, 
X-ray emission spectroscopy, and M\"{o}ssbauer spectroscopy. 
$L_3$-edge X-ray absorption spectroscopy on the rare-earth ions is very suitable 
for experiments at high pressures \cite{rueff1} and/or 
high magnetic fields \cite{yhm07, yhm09}, 
because high-energy X-rays have strong penetrating power, so an ultrahigh-vacuum condition is not necessary.
Determination of the valence state can be done in very high magnetic fields of up to 40~T by X-ray absorption measurement \cite{yhm07, yhm09}.

In the present study, we have studied the temperature dependence of Yb valence in $\beta$-YbAlB$_4$ \cite{nakatsuji08} and 
$\alpha$-YbAlB$_4$ \cite{matsumoto} by 
using $L_3$-edge x-ray absorption and emission spectroscopy.
The X-ray absorption spectra were measured in high magnetic fields up to 32~T and 
a small, but finite valence increase was 
found at 40~K. 

\section{EXPERIMENT}

The experiment was done at the beamline BL39XU of SPring-8, Japan. 
Pulsed high magnetic fields were generated by using a miniature magnet \cite{yhm07}.   
$\beta$- and $\alpha$-YbAlB$_4$ crystals were powdered and mixed with epoxy resin so that the effective 
thickness was appropriate for obtaining an X-ray absorption intensity of $\mu t \sim 1$, where $\mu$ and $t$ 
are the absorption coefficient and the thickness of the sample, respectively.
A single crystal of $\beta$-YbAlB$_4$ was also used for X-ray absorption and X-ray emission spectroscopy. 
The details of the experimental techniques using a pulsed high field are described elsewhere \cite{yhm07, yhm09}.

\section{RESULTS and DISCUSSION}
Figure 1 shows the X-ray absorption spectrum at the $L_3$-edge (from the $2p$ to $5d$ transition) 
in a powdered $\beta$-YbAlB$_4$ at 10~K. 
The spectrum of a single crystal of $\beta$-YbAlB$_4$ was also measured and found to be very similar to that of the 
powdered sample.
The solid curve is the result of the curve-fitting analysis. 
We analyzed the X-ray absorption spectra (XAS) by using a fitting function that consists of a Lorentzian and an arctangent function. 
This is a standard method for analyzing the absorption edge and has previously been used in an X-ray absorption 
study of YbInCu$_4$ \cite{yhm07}. 
The dashed and dot-dashed curves are the results of the curve-fitting 
for Yb$^{3+}$ and  Yb$^{2+}$ states, respectively.
Background is represented by a dotted line. 

The main absorption peak around 8.950~keV due to the Yb$^{3+}$ state is distinct.
On the other hand, the absorption band around 8.945~keV due to the Yb$^{2+}$ state is not clearly seen in the 
spectrum.
Since the valence ($v$)  of $\beta$-YbAlB$_4$ was reported to be 2.75 at 20~K \cite{ohkawa}, about a 25\% spectrum weight 
is expected to be observed.
The lifetime broadening effect of the $2p$ core hole can be the reason the Yb$^{2+}$ band at the low energy is not 
very clear.
However, as we show later, the Yb$^{2+}$  band is confirmed to be located around 8.945~keV by the 
magnetic field dependence of XAS.

Although the $v$ is deduced to be 2.84 from the fitting results shown in Fig.~1, the error bar can be as large as 0.05 
because of the broadness of the XAS.
\begin{figure}
\includegraphics[width=10.0cm]{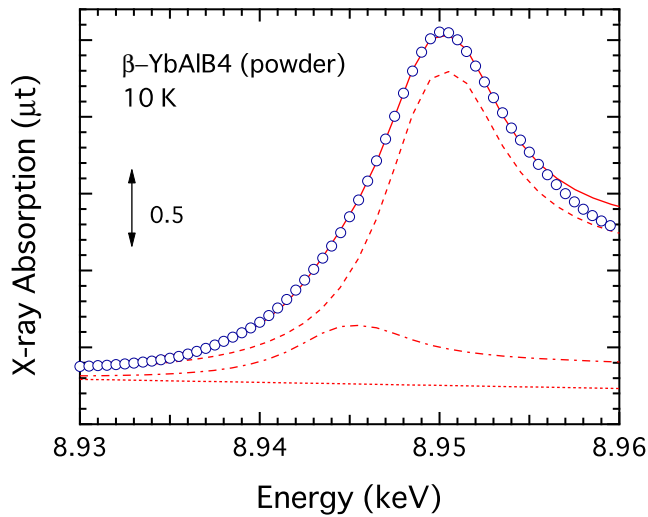}
\caption{(Color online) X-ray absorption spectrum of $\beta$-YbAlB$_4$. 
The solid curve is the result of the curve-fitting analysis. 
The dashed and dot-dashed curves are the results of the curve-fitting 
for the Yb$^{3+}$ and Yb$^{2+}$ states, respectively.
Background is represented by a dotted line.
}\label{fig1}
\end{figure}
Therefore, we determined $v$ at low temperatures by another method, i.e.,  X-ray emission spectroscopy 
that enables us to observe the XAS 
with smaller broadening effect of the core hole \cite{hayashi}.  

The X-ray emission spectra (XES) with different excitation energies at 2 and 280~K are shown in Fig.~2. 
It is found that the emission band due to Yb$^{2+}$ state is larger at  2~K compared to that at 280~K, indicating that the valence decreases 
with decreasing temperature. 
\begin{figure}
\includegraphics[width=8.5cm]{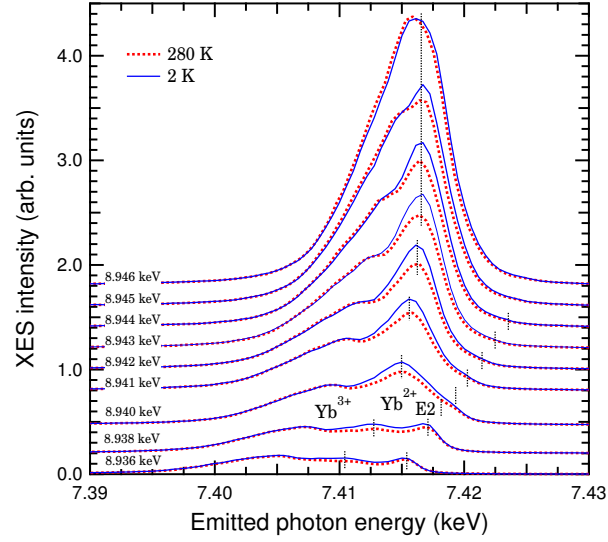}
\caption{ (Color online)
Yb $L\alpha_1$ X-ray emission spectra at different excitation energies in a $\beta$-YbAlB$_4$ single crystal at 2 and 280~K.} 
\label{fig2}
\end{figure}
From the analysis of the XES $v$ was found to be $2.78 \pm 0.01$ and $2.84 \pm 0.01$ at 2 and 280~K, respectively.  \cite{kanai} 
Since the optical transition process of XES is more complicated than that of XAS, the valence state cannot be directly obtained by using 
the spectral weight. The detailed analysis is explained in elsewhere \cite{kanai} and is beyond the scope of this paper. 

It is found that the XES show distinct emission bands due to the Yb$^{2+}$ and Yb$^{3+}$ contributions.
This is because 
the lifetime of the $3d$ core hole that exists in the final state of the emission process (from the 3$d$ to 2$p$ $L\alpha_1$ transition) 
is expected to be much longer than that of the $2p$ core hole.
This high-resolution spectroscopy is more reliable than XAS when the XAS is broad, making it difficult to distinguish each valence 
state separately.
Hence, we take $2.78 \pm 0.01$ for the valence value of $\beta$-YbAlB$_4$ at 2~K, and the temperature dependence of $v$ 
obtained from XAS is used as the relative change.
The relative change in the spectrum is determined precisely, and the valence change is determined with much smaller error 
bar compared to the error bar for determination of the absolute value.
The temperature variation of $v$ is shown in Fig.~3(a).
\begin{figure}
\includegraphics[width=13cm]{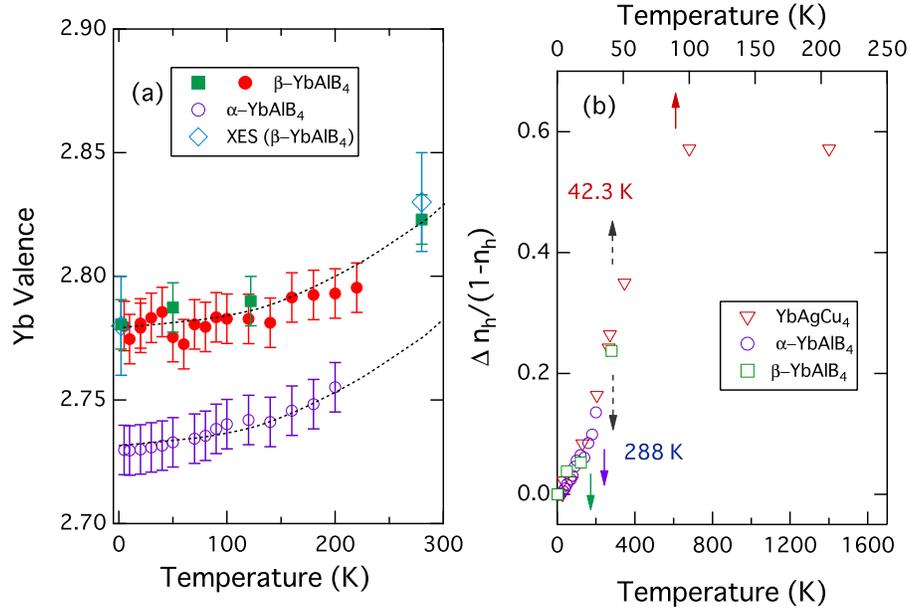} 
\caption{ (Color online)
(a) Temperature dependence of the Yb valence in $\beta$-YbAlB$_4$ and $\alpha$-YbAlB$_4$. The closed squares are the results 
of the single crystal of $\beta$-YbAlB$_4$, and closed and open circles are the results from the powdered $\beta$- YbAlB$_4$ and 
$\alpha$-YbAlB$_4$ samples, respectively.  
Open diamonds denote the results of XES in the single crystal of $\beta$-YbAlB$_4$. 
The dashed curves are guides for the eyes. 
(b)  Temperature dependences of the Yb valence in the single crystal $\beta$-YbAlB$_4$ (open squares) 
and powdered $\alpha$-YbAlB$_4$ (open circles) are shown together with that of the Yb valence in YbAgCu$_4$ (open triangles)  \cite{yhm12}.
}
\label{fig3}
\end{figure}
$v$ of $\alpha$-YbAlB$_4$ is also plotted as a function of temperature.
Because we did not measure XES of $\alpha$-YbAlB$_4$, we used the literature value at 20~K (2.73) \cite{ohkawa} for the 
$v$ value at 5~K, and the relative change in $v$ with increasing temperature was deduced by the change in the XAS in this work.
It is found that the valence gradually increases with increasing temperature, and the increasing trends for 
 $\beta$- and $\alpha$-YbAlB$_4$ look similar.

The occupation number of the hole in the 4$f$-shell ($n_h$) is deduced from $v-2 $. 
According to the G\"{u}tzwiller approximation, the deviation of $n_h$ from unity ($1-n_h$) can be expressed by 
$1-n_h= \frac{kT_F^*}{\Gamma}$, where $kT_F^*$ and $\Gamma$ correspond to the kinetic energy of the $f$ electrons with the 
electron interaction and that without the interaction, respectively \cite{heavy}.
An itinerant character manifests itself in the deviation of the occupation number of the $f$ hole from unity $1-n_h$.
%
%
If we define $\Delta  n_h$ as the difference of 
$n_h$ from that at the lowest temperature, the value 
$\Delta  n_h$/($1- n_h$) corresponds to the relative change in the degree of itinerancy 
when the temperature is increased.

In Fig.~3(b), the temperature dependences of $\Delta  n_h$/($1- n_h$) in single crystal $\beta$-YbAlB$_4$ (open squares) 
and powdered $\alpha$-YbAlB$_4$ (open circles) are shown.
$\Delta  n_h$/($1- n_h$) in another heavy fermion compound YbAgCu$_4$ \cite{yhm12} is also plotted (open triangles).  
YbAgCu$_4$ is one of the materials that are expected to be located in the vicinity of the QCP of the valence transition \cite{wata09}. 
The scale of the horizontal axis is set differently for YbAlB$_4$  (bottom) and YbAgCu$_4$ (top) so that the temperature variation of 
$\Delta  n_h$/($1- n_h$)  in both materials almost overlap.

The valence transition temperature ($T_v$) was determined to be 42.3~K in YbAgCu$_4$ in our previous work \cite{yhm12, nakamura12}.  
If we simply assume that the temperature dependence of the valence in YbAlB$_4$  is scaled to that in YbAgCu$_4$, the hypothetical valence transition temperature ($T_v^*$)
is 288~K for YbAlB$_4$.
This value seems to be consistent with the fact that $\beta$-and $\alpha$-YbAlB$_4$ have a large characteristic temperature $T_0$ $\sim$  250~K  below which the magnetic moments are screened to form the HF state \cite{matsumoto}.
$v$ becomes about 2.96 and the local magnetic moments are restored in YbAgCu$_4$ at about 70~K \cite{yhm12, nakamura12}.
According to Fig.~3(b), the localized magnetic moments are expected to be restored in $\beta -$ and $\alpha -$YbAlB$_4$ at corresponding 
temperature around 480~K.  

Since the characteristic temperature $T_0$ of YbAlB$_4$ is as high as 250~K \cite{matsumoto},   a strong magnetic field is necessary to 
see the effect of the magnetic field on the valence state. 
Figure 4 shows XAS at 0 and 32~T in the lower panel and the difference spectrum between 0 and 32~T (dXAS) in the upper panel. 
The measurement temperature was 40~K.
The XAS seem to be almost identical at 0 and 32~T.
However, the positive peak around 8.950~keV and negative peak around 8.945~keV in dXAS are clear evidence of an increase in the 
Yb$^{3+}$ component and a decrease in the Yb$^{2+}$ component, respectively.
The solid curve is the result of the curve fitting.
The agreement of the fitting and the experimental data confirms that a small valence increase, $\Delta v \sim 0.002$, is induced by a magnetic 
field of 32 T. 
Although a similar measurement was made at 5~K for the same sample no clear valence change was observed.
Therefore, the observed field-induced small valence change was assisted by a thermal excitation at 40~K. 

\begin{figure}
\includegraphics[width=10cm]{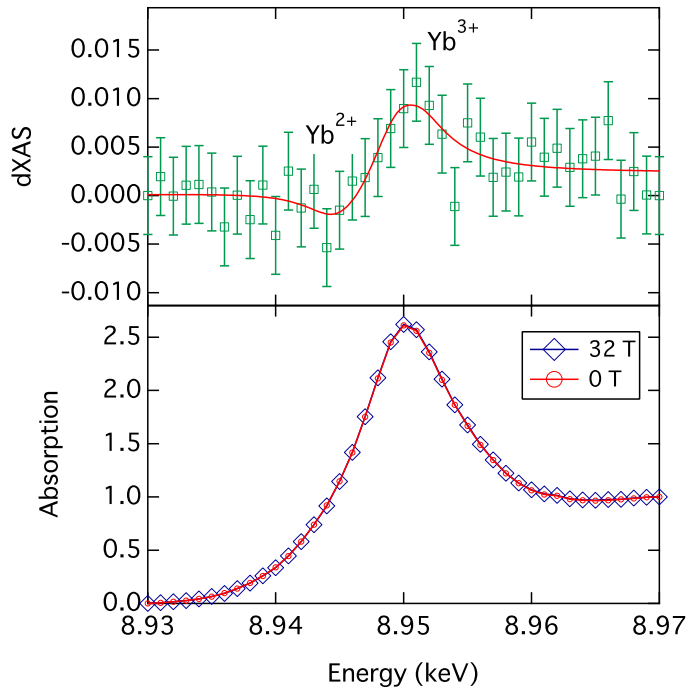}
\caption{ (Color online)
X-ray absorption spectra (XAS) in a powdered $\beta$-YbAlB$_4$ sample at 0 and 32 T (lower panel). 
The difference spectrum between the XAS at 0 and 32 T is shown in the upper panel, together with the result 
of the curve fitting. A magnetic field is applied parallel to the c-axes of the powders that were aligned by means of a steady magnetic 
field of 14~T when the epoxy resin was solidified.
 The measurement temperature was 40 K. }
\label{fig4}
\end{figure}

\section{SUMMARY}

XAS and XES in $\beta$- and $\alpha$-YbAlB$_4$ were investigated at several temperatures and a high magnetic field of 32~T. 
The observed Yb valence is 2.78 $\pm$ 0.01 in $\beta$-YbAlB$_4$ at 2 K by using XES. 
The Yb valences of these materials are found to increase with increasing temperature and by applying a magnetic field.
By comparison with another heavy fermion compound YbAgCu$_4$, the hypothetical valence transition temperature in 
$\beta$- and $\alpha$-YbAlB$_4$ is expected to be about 288 K; this value is close to the characteristic temperature $T_0$ of  
these materials as heavy fermion systems. 
Experiments at fields above 100 T are required to observe more distinct field-induced valence changes. 
Magnetization experiments in the 100-T range by means of a single-turn coil technique \cite{takeyama} are currently ongoing at 
the Institute for Solid State Physics, University of Tokyo. \\

We thank H. Toyokawa and T. Uruga of Japan Synchrotron Radiation Research Institute/SPring-8 for technical support of 
the PILATUS detector. 
For the analysis of the XES results, we thank N. Kanai and H. Hayashi of Japan Women's University. 
This work was partly supported by Grants-in-Aid for Scientific Research B (22340091), Scientific Research A (22244047), 
Scientific Research C (24540389) and Young Scientists B (22740208) provided by the Ministry of Education, 
Culture, Sports, Science and Technology (MEXT), Japan.
The synchrotron radiation experiments were performed with the approval of the Japan Synchrotron Radiation 
Research Institute (JASRI) (Proposal No. 2011B1090, 2011B2097 and 2012A1173).

\end{document}